\documentclass[final,12pt]{elsarticle}

\usepackage{amssymb}
\usepackage{amsmath}
\usepackage{siunitx}
\usepackage{booktabs}
\usepackage{color}
\usepackage{subfigure}
\usepackage{natbib}
\usepackage{url}
\usepackage{doipubmed}
\usepackage{graphicx} 
\usepackage[UKenglish]{babel}

\journal{ArXiv}

\begin{document}

\begin{frontmatter}



\title{Towards Gigayear Storage Using a Silicon-Nitride/Tungsten Based Medium}


\author{Jeroen de Vries$^{1,\star}$, Dimitri Schellenberg$^{1}$, Leon Abelmann$^{1,3}$, Andreas Manz$^{2,3}$ and Miko Elwenspoek$^{1,2}$}

\address{%
$^{1}$ MESA+ Institute for Nanotechnology, University of Twente, Drienerlolaan 5, Enschede, The Netherlands\\
$^{2}$ Freiburg Institute for Advanced Studies (FRIAS),
Albert-Ludwigs-Universit\"at Freiburg, Albertstra\ss{}e 19, D-79104,
Freiburg i.Br, Germany\\
$^3$ KIST-Europe, Campus E7, D-66123, Saarbr\"ucken, Germany}

\let\thefootnote\relax\footnote{$^\star$ Corresponding author: j.devries-5@alumnus.utwente.nl}


\begin{abstract}
Current digital data storage systems are able to store huge amounts of data. 
Even though the data density of digital information storage has increased tremendously over the last few
decades, the data longevity is limited to only a few decades. 

If we want to preserve anything about the human race which can outlast the human race 
itself, we require a data storage medium designed to last for 1 million to 1 billion years.
In this paper a medium is investigated consisting of tungsten encapsulated by siliconnitride 
which, according to elevated temperature tests, will last for well over the suggested time. 
\end{abstract}


\end{frontmatter}


\section{Introduction}
\label{S:1}

\begin{figure}[t]
  \centering
  \includegraphics[width=.6\columnwidth]{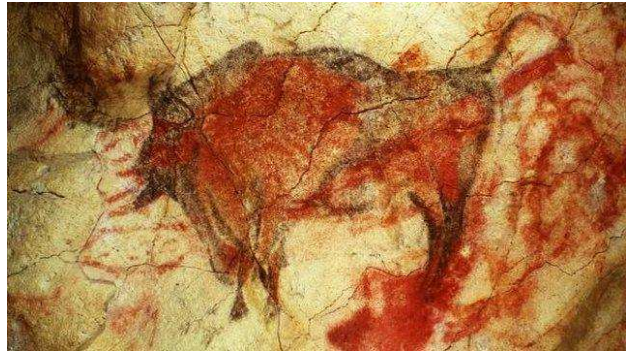}
  \caption{Cave painting from the El Castillo cave in Spain, 
  estimated to be over \num{40000} years old}
  \label{fig:ElCastillo}
\end{figure}

The human race has achieved many things we consider worth
storing. From paintings found in caves as shown in figure \ref{fig:ElCastillo} 
to pieces currently 
on display in musea. Whether it is music, art, 
literature or scientific breakthroughs, people have 
tried to ensure preservation of information for future generations.

Musea are filled with art and most of the music and movies today are
accessible through the internet, but for how long? At some point
humanity as we know it will cease to 
exist~\citep{Elwenspoek2011} and
slowly all our achievements will disappear. Given sufficient
time, all memory of humanity will be erased. 

To ensure that knowledge about human life is available
for many future generations or even future lifeforms 
we require a form of data storage suitable
for storage at extreme timescales. 

There are of course some requirements for such 
a data storage system. The system should be able 
to survive for at least 
the required time without losing its content. 
The data should be easily decodable and the
data carriers should be stored in locations
likely to change little over 1 million years to
ensure that the data carrier does not end up
at the bottom of the ocean after a million years.
It is also necessary of course to determine
what data is relevant to store. All aspects of such a storage 
system are combined in a 
multidisciplinary project called 
``the human document project''~\citep{Manz2010}.
 
Although each of these aspects is of vital 
importance to ensure that the stored data
will survive and is recognised as such, in  
this paper we will only focus on the fabrication 
of a data carrier which can survive 
for over one million years.

To preserve information for future generations, people
have stored data in various ways. In recent years the 
storage capacity has increased tremendously due to the 
possibilities created by digital information storage. Where in 
1956 the IBM 305 RAMAC was capable of storing \SI{5}{\mega B} of data
using fifty 24'' diameter disks, currently \SI{4}{\tera B} can be 
stored on four 3.5'' diameter disks. This means a
decrease of form factor and power consumption while greatly
increasing the storage capacity. This increase in capacity
has not yet reached its limit~\citep{Muraoka2011}.  Although there has
been a huge increase in storage density, the data 
longevity is limited to about a decade.

Storage systems like DVD's will not last much longer and 
tape storage will (if stored in a proper environment)
last only several decades. Archival paper is expected to last 
up to \num{500} years but only if it is stored in a suitable environment.

A recent long-term data storage system is created
by the 'Long Now Foundation' where the information is 
etched into a substrate and then electroformed in solid 
nickel~\citep{LongNow2013}. The resulting disk contains \num{13000} pages
of information on more than \num{1500} human languages. 
A disadvantage of the disk is that its surface is fragile
and can easily be scratched so an encasing for the 
disk is essential for long time survival and repeated
readout might damage the surface of the disk. 


If we want to store information for much longer timescales 
none of the above described media will be suitable and
a new type of storage medium is required where the longevity 
of the data is more important than the storage density.

In order to store information for very long timescales, a medium is
required which will survive for at least this period and still contain
discernible data. We believe that the relevant storage time should be 
at least 1 million years and at most 1 billion years~\citep{Elwenspoek2011}.

The type of storage medium suitable for these timescales 
should likely be fabricated especially for this project. 
There are many types of media which will be suitable, of
which the most exotic is DNA based data storage within a living organism which 
reproduces itself~\citep{Petr2011}. We have chosen a disk based storage system 
because such a system can be created with technology 
which is already available and will likely be easier recognisable  
than a data carrier inside a living organism. Although
magnetic based data storage is highly suitable for current 
purposes, its energy barrier is too low to survive for 1 million
years. Moreover, the data could be erased or modified by strong magnetic fields. 

We expect that the data that needs to be stored for the Human Document Project will have been
thoroughly verified and is not subject to change. Therefore a
'write-once-read-many' (WORM) type data system would be sufficient.

Readback of the disk should be possible using electromagnetic waves
which, we expect, any sufficiently developed society would be able to use. 
The disk can contain multiple levels of data with different data densities. Using
visible light the low density data could be read by eye or using optical
microscopy. Higher density data could be made visible by for instance
electron beams. A dedicated readback system would in this case not be
stored with the data carrier because such a system would also need to survive for at least
as long as the disk. Plans to create a readback system could be part
of the information on the data carrier.

\section{Theory}

All data is volatile so it is not possible to store data
indefinitely. Even data engraved in a marble slab will eventually
erode away. The longevity of a data carrier can of course be
increased by storing it in a monitored environment such as in
current paper archives where the humidity, temperature, pressure
etc. are controlled. This however, would require 
that the environment also survives and is maintained for 
at least as long as the data carrier.
We believe that the chances of such an environment surviving for a
million years in operating conditions are slim, so we prefer a data carrier which has a high
chance of surviving without such a dedicated environment.

In order to create a stable data carrier, able to preserve 
data for a million years, a high energy barrier 
against erasure is required. This thermal stability of 
data is a well studied aspect in magnetic data storage~\citep{Charap1997}. 

\begin{figure}
  \centering
  \def\svgwidth{0.5\columnwidth}
  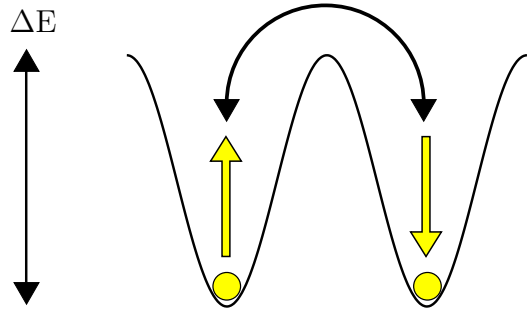
  \caption{Data is stored in the state of a system, which can be in
    two or more energy minima.}
  \label{fig:superparamagnetic-reversal}
\end{figure}

Using a simple but effective theory based on the Arrhenius law, as
often used by scientists in the hard disk industry, we can determine
the energy barrier required for a certain storage time. For the data
storage system we assume that one bit of data is stored in a way
  much like in current magnetic
  storage systems. The data is stored in one of the energy minima of
the system, which is separated from the other minima by an energy
barrier $\Delta E$, as indicated in figure \ref{fig:superparamagnetic-reversal}. 
If the system would be
kept at a temperature of \SI{0}{\kelvin}, there would be no thermal
fluctuations and the system would stay in this state indefinitely.
However at elevated temperatures, the probability that the system will
jump to another energy minimum after time $t$, $P_\text{sw}$ is given by the
Arrhenius law~\citep{Wernsdorfer1997}.

\begin{align}
P_\text{sw} &= 1-\exp\left(-t/\tau\left(T\right)\right)\\
\tau\left(T\right) &= f_0^{-1} \exp\left(\Delta E/\text{k}_\text{B}T\right)
\label{eq:Arrhenius}
\end{align}

Where $\tau$ is the decay time [\si{\second}], k$_\text{B}$ Boltzmann constant [\si{\joule\per\kelvin}],
$T$ the absolute temperature [\si{\kelvin}] and $f_0$ the attempt frequency
[\si{\hertz}] which is related to the atomic vibrations
and is in the order of \SI{1}{\giga\hertz} for 
magnetic particles~\citep{Weller1999}.

We assume that the probability of switching  from
one of the energy minima to another is low, so we can neglect
secondary processes like switching back to the original
energy minimum. We also assume that the switching behaviour
of one element does not influence the switching behaviour
of the others. In this case the number of incorrect 
bits of information in a large data set of $N$ bits is given
by $P_\text{sw}N$. In modern storage systems error fractions up to
$\alpha$ of \num{1E-5} can be comfortably corrected by suitable 
error codes. Rewriting equation \ref{eq:Arrhenius} gives:

\begin{align}
\tau &> -t/ \text{ln} \left(1-\alpha\right) \approx t/\alpha \hspace{1cm} \text{for}\ \alpha\ll 1\\ 
\text{and} \hspace{1cm} \Delta E/\text{k}_\text{B}T &> \text{ln}\left(t f_0/\alpha\right)
\label{eq:tau}
\end{align}
 
For a data storage time of 1 million years with $\alpha$=\num{1E-5} and $f_0$=\SI{1}{\giga\hertz}, 
$\Delta$E should be \num{63}~k$_\text{B}T$, for 1 billion years the energy barrier 
should be raised to \num{70}~k$_\text{B}T$ (\SI{1.8}{\electronvolt} at room temperature). These energy barriers are given for the ideal case, where there are no distributions of properties which lower the energy barriers. These 
values are well within the range of today's technology. 

To prove that the data will in fact remain without uncorrectable 
errors for one million years is another challenge. To wait for a million years to be certain that the data 
remains would be slightly impracticable, so an accelerated ageing test
is required. Starting from equation \ref{eq:tau} there are three variables we can act upon, 
the testing time $t_t$, the observed number of errors during the test
$\alpha_t$ and the temperature at which the tests are performed $T_t$.
\begin{align}
  \Delta E/\text{k}_\text{B}T_\text{t}>\text{ln}(t_\text{t}f_0/\alpha_\text{t})
  \label{eq:HDP3}
\end{align}

By observing many bits, we can determine error rates lower 
than \num{1E-5} and extrapolate from there when this value 
will be reached. By testing at higher temperatures, we can 
increase the number of errors per time unit. Combining equation \ref{eq:tau} 
and equation \ref{eq:HDP3}, the temperature at 
which the test is performed is given by:

\begin{align}
  \label{eq:4}
  T_t > 
  T \frac{\ln\left(\frac{t
        f_0}{\alpha}\right)}{\ln\left(\frac{t_\text{t} f_0}{\alpha_t}\right)}
\end{align}

Taking for instance an observed error rate ten times better 
than the desired rate (so $\alpha_t$=\num{1E-6}), the required 
testing temperature to prove that the data is stable for a 
million years within a year is \SI{380}{\kelvin}. Table \ref{tab:testingtemperatures} 
lists values for different testing and storing times, which are well 
within the experimental range.

\begin{table}
\centering
\begin{tabular}{@{}lllr@{}} 
	\toprule
		Storage period & 1 hour & 1 week & 1 year\\
	 \midrule
		\num{1E6} years &  \SI{461}{\kelvin} & \SI{411}{\kelvin} & \SI{380}{\kelvin} \\
		\num{1E9} years &  \SI{509}{\kelvin} & \SI{455}{\kelvin} & \SI{420}{\kelvin}\\ \bottomrule
\end{tabular}
  \caption{Testing at elevated temperatures to prove data retention 
  for different timescales at
    T=\SI{300}{\kelvin} with $\alpha$=\num{1E-6}}
  \label{tab:testingtemperatures}
\end{table}

This simple model is a first step towards proving
that data will be retained for at least one million years by means of an
elevated temperature test. This method is also known as accelerated
ageing~\citep{Zou1996}. The method does have its shortcomings however.

\subsection{Attempt frequency} The simple Arrhenius model assumes
that the attempt frequency to overcome the energy barrier is much higher than the
reciprocal of the testing time (in other words, there should be many
attempts to switch the entire bit within the testing period). This
condition is most likely met. Other causes of data loss, such as
theft, meteor impact or the sun entering the red giant
phase~\citep{Elwenspoek2011} cannot be revealed by accelerated ageing
(fortunately).

\subsection{Local minima} The model assumes that one single event
switches the bit. Whereas this might be true for patterned magnetic
disks, it certainly is not the case for all data storage systems. In
the simple model, we assume that the system only possesses global
minima, where local minima might exist which can serve as intermediate
steps towards overcoming the energy barrier between global minima
states. An example of such a system is the phase change medium (used
in rewritable optical discs) where data is stored in the position of
atoms, which can reside in a huge number of local minima. In this case
the chance of switching should be calculated as a cascade of events,
each possibly with their own attempt frequency. The process could then
be described by an Arrhenius cascade~\citep{Bardou2000}.

\subsection{Temperature dependence} The simple model assumes that
the energy barrier is independent of the temperature. In most situations
this will not be the case. In general the energy barrier will
decrease with increasing temperature. An accelerated test will then underestimate the 
lifetime of the medium. The model also assumes that attempt frequencies are independent 
of the temperature. Fortunately, the attempt frequency appears in both
the numerator and denominator of equation \ref{eq:4}. A ten-fold
increase in attempt frequency only reduces the temperatures in table \ref{tab:testingtemperatures} by \SI{10}{\kelvin}. Detailed knowledge on the
attempt frequency is of lesser importance, as is obvious from equation \ref{eq:tau}.

Despite the shortcomings of the single switch Arrhenius model, it is
still of interest to perform actual accelerated tests on a real medium,
especially at temperatures much higher than listed in table \ref{tab:testingtemperatures}. 

\section{Fabrication}

\begin{figure}
  \centering  
  \def\svgwidth{0.5\columnwidth}
  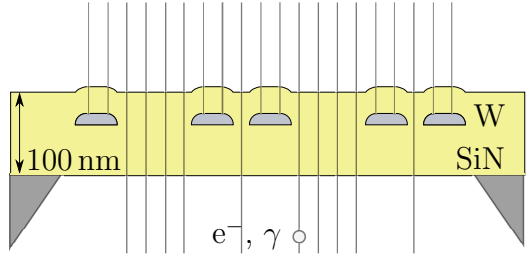
	\caption{W-$\text{Si}_\text{3}\text{N}_\text{4}$ WORM medium, which is transparent to electron or photon beams}
  \label{fig:WSiNMedium}
\end{figure} 
The suggested data storage system consists of a medium where the data is 
represented by one material embedded within a second, different material as
schematically described in figure \ref{fig:WSiNMedium}.  We have
selected as the base materials tungsten for the data and
Si$_3$N$_4$ for the encapsulating material. Tungsten has a high
melting temperature and high activation energy, furthermore it has a
low thermal expansion coefficent. The Si$_3$N$_4$ has a high fracture toughness
and low thermal expansion coefficient. Another important feature of the
Si$_3$N$_4$ is its transparency to light. A very thin film would also be
transparent to electron beams.  These materials are readily available
and generally used in microfabrication.
\begin{figure}
  \centering  
  \def\svgwidth{0.5\columnwidth}
  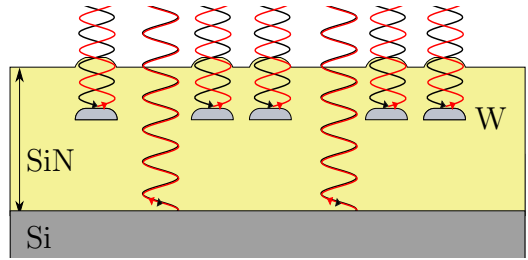
	\caption{W-$\text{Si}_\text{3}\text{N}_\text{4}$ WORM medium, which is readable
	by constructive interference on the silicon base and destructive 
	interference on the tungsten islands.}
  \label{fig:WSiNMediumRefl}
\end{figure}

An alternative to the transparent medium is a medium with high
contrast in reflection. Contrast can be enhanced by using an optical beam 
of a single wavelength. If the layer thickness is tuned correctly, 
constructive interference can occur in the parts where there is no metal and
destructive interference in the parts where metal is present or vice versa, as shown 
schematically in figure \ref{fig:WSiNMediumRefl}.
This makes it possible to have a much thicker base because the 
sample does not need to be optically transparent. It should 
however still be ensured that not all the light gets absorbed by the nitride layer.
With light in the visible spectrum, this method can be used for low density data. 

\subsection{Optical readable data}
As a demonstration, data is written in two-dimensional (matrix)
barcodes, which can be read back by a camera and computer. These
two-dimensional barcodes were introduced for cases where more
information needs to be stored than can be accomodated by their one-dimensional
predecessors, but are now becoming increasingly popular. The
implementation we chose was the quick response (QR) code~\citep{Vongpradhip2013}, which can
be easily decoded by todays smartphones.  The level of QR code
containing the largest amount of information can lose up to
\SI{7}{\percent} of the data before the code becomes unreadable.  

\begin{figure}[t]
  \centering
  \def\svgwidth{0.45\columnwidth}
   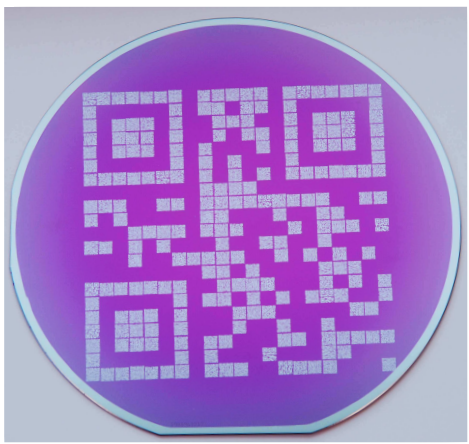
  \caption{Design of the HDP disk. Each pixel in the large
  QR codes consists of a QR code.}
  \label{fig:HDPDisk}
\end{figure}

For the
encoding of the final disk, it is likely that a coding scheme would be
required which focusses on easy decodability. By keeping the size of
the QR code low, it is possible to read out the disk by an optical
microscope. For the demonstration, the entire disk was covered with a
centimeter sized QR code. Each pixel of the code consists of a set of
much smaller QR codes with pixels of only a few micrometers in size as shown in figure \ref{fig:HDPDisk}.

For design of the thickness of the nitride layers and the tungsten
layers, optical calculation software was used
in order to determine the optimum thickness for maximum
extinction and amplification in the layers. The wavelength was
chosen to be \SI{550}{\nano\meter}.

The medium consists of a \SI{338}{\nano\meter} layer of 
LPCVD $\text{Si}_3\text{N}_4$ on a bare silicon 
wafer. The tungsten is patterned using optical lithography
and a mask containing the QR codes.
The pattern is etched using Ar ion beam etching and 
a top layer of PECVD nitride of \SI{225}{\nano\meter} is deposited 
on top of the tungsten patterns. The process steps 
are schematically shown in figure \ref{fig:SampleDiagramLongLIL}
where the silicon removal in step 8 is not necessary.

\subsection{Line patterns}
Because readout by optical microscope means that the data
density is low, it is also necessary to have a higher
density storage method. The higher data density
storage can be achieved by embedding the data in the nitride
for readout by electron beam. Here we assume that the medium
should be transparent to electrons, which means that the disk
will be very thin and fragile. 

For the high data density sample, tungsten lines are used instead of islands. 
With this sample it is possible to simulate high density data with a linewidth below 
\SI{100}{\nano\meter}.

\begin{figure}[t]
  \centering
  \def\svgwidth{0.9\columnwidth}
   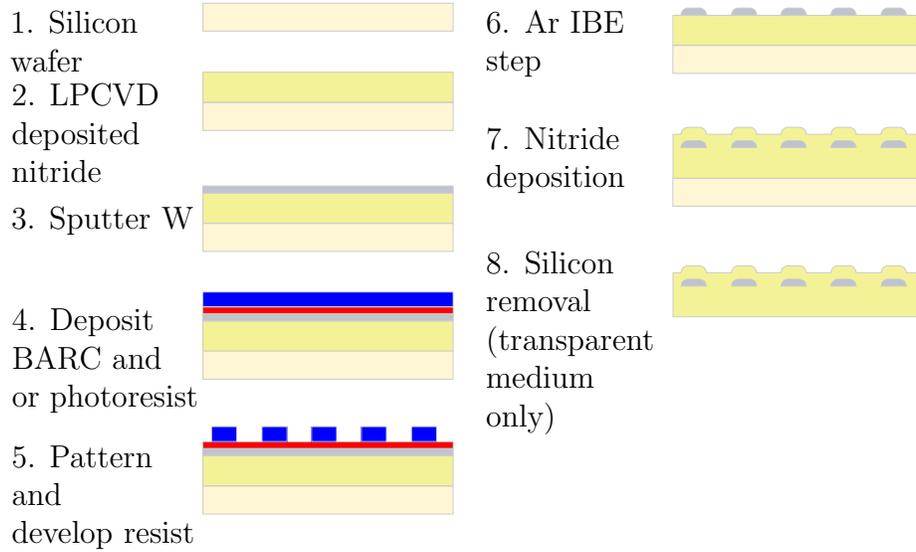
  \caption{Design of the HDP disk. Each pixel in the large
  QR codes consists of a QR code.}
  \label{fig:SampleDiagramLongLIL}
\end{figure}

The lines will make it easier to create very small structures and inspect 
the sample after thermal exposure by means of an SEM. 
A drawback of these lines is the variation in stress in the 
sample in the direction along the lines compared to across the lines. 
The process steps are described in figure \ref{fig:SampleDiagramLongLIL}.

The test sample is created by depositing a layer of 
\SI{230}{nm} silicon nitride on a cleaned bare silicon wafer by 
an LPCVD process. On top of the silicon a layer of \SI{50}{nm} 
tungsten is deposited by magnetron sputtering. 
A layer of DUV 30-J8 bottom anti-reflective coating (BARC) is 
applied by spin coating,  which limits the standing waves 
in the resist and improves the vertical sidewalls. 
On top of the BARC, a layer of PEK-500 positive resist is spun.

\begin{figure}[t]
  \centering
  \subfigure{
  \def\svgwidth{0.45\columnwidth}
   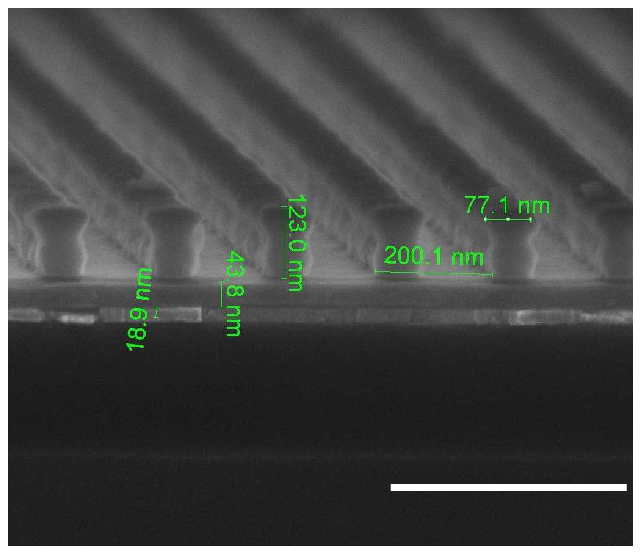
   }
	\hskip1mm
	\subfigure{
	\def\svgwidth{0.45\columnwidth}
  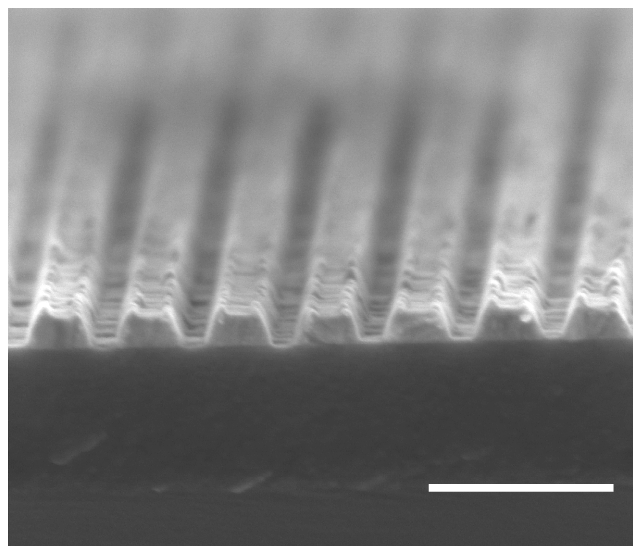
   }
 	\vskip1mm
	\subfigure{
	\def\svgwidth{0.45\columnwidth}
  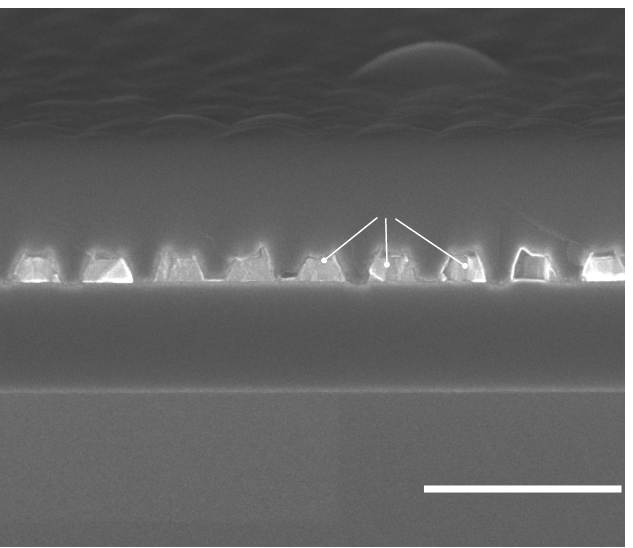
   }
   \hskip1mm
	\subfigure{
	\def\svgwidth{0.45\columnwidth}
  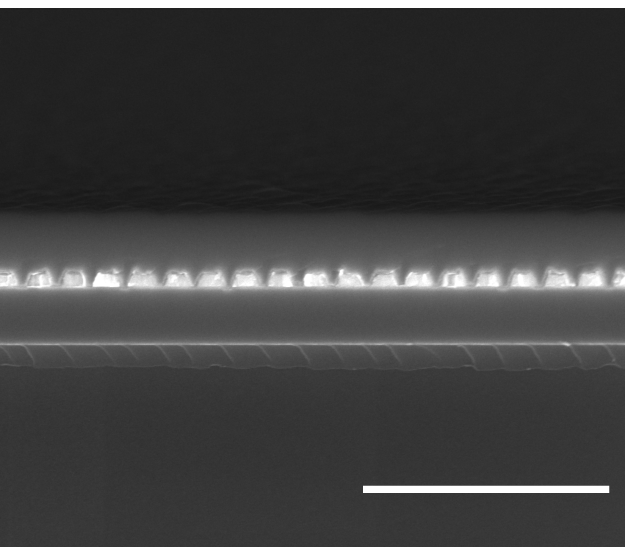
   }

  \caption{a) Scanning electron micrograph of the test sample before
    etching. b) Scanning electron micrograph of the test sample
    after etching containing W lines. c) Scanning electron micrograph of the cross-section of the encapsulated lines in the test sample
    d) Scanning electron micrograph of the sample after 1 hour at \SI{473}{\kelvin}}
  \label{fig:SEMimages}
\end{figure}

Laser interference lithography (LIL) is used to create the pattern
of \SI{100}{nm} wide lines~\citep{Luttge2007}. For the actual data carrier
either a standard lithography mask process could be used or a
laser could be used to write the data in the resist. The resist 
is developed in OPD4262 after exposure. The process flow is depicted 
in figure \ref{fig:SampleDiagramLongLIL} and a scanning
electron micrograph of the developed sample is shown in figure \ref{fig:SEMimages}~a.

A short O$_2$ reactive ion beam etching (RIBE) step is used to
transfer the pattern into the BARC layer. The BARC pattern is 
transferred into the tungsten layer by argon ion beam milling as can
be seen in figure \ref{fig:SEMimages}~b. 

The entire sample is subsequently covered with $\text{Si}_3\text{N}_4$ by a
PECVD process to encapsulate the tungsten lines. The result is shown in figure \ref{fig:SEMimages}~c. 
In the cross-section image
from bottom to top, the silicon, the LPCVD silicon-nitride, the 
tungsten lines and the PECVD silicon nitride can be seen. 
The $\text{Si}_3\text{N}_4$ in the final product is much thicker 
than the thickness schematically shown in figure \ref{fig:WSiNMedium} 
to observe possible spreading of the tungsten clearly. 

For the optical transparent sample, the silicon needs to be removed 
from the bottom of the sample and a 
medium with tungsten lines encapsulated in a $\text{Si}_3\text{N}_4$ matrix 
remains. The silicon removal of step 8 shown in figure \ref{fig:SampleDiagramLongLIL}
is not performed to ensure mechanical stability of the sample. 

\section{Elevated temperature test}

By testing the sample at relevant temperatures it can be shown that it
should in principle be possible to store data for at least one million
years. A second interesting test would be to investigate whether the
sample would survive higher temperatures which would for instance 
occur during a house fire.

\subsection{Optical readable data}

\begin{figure}[t]
  \centering
  \subfigure{
  \def\svgwidth{0.25\columnwidth}
   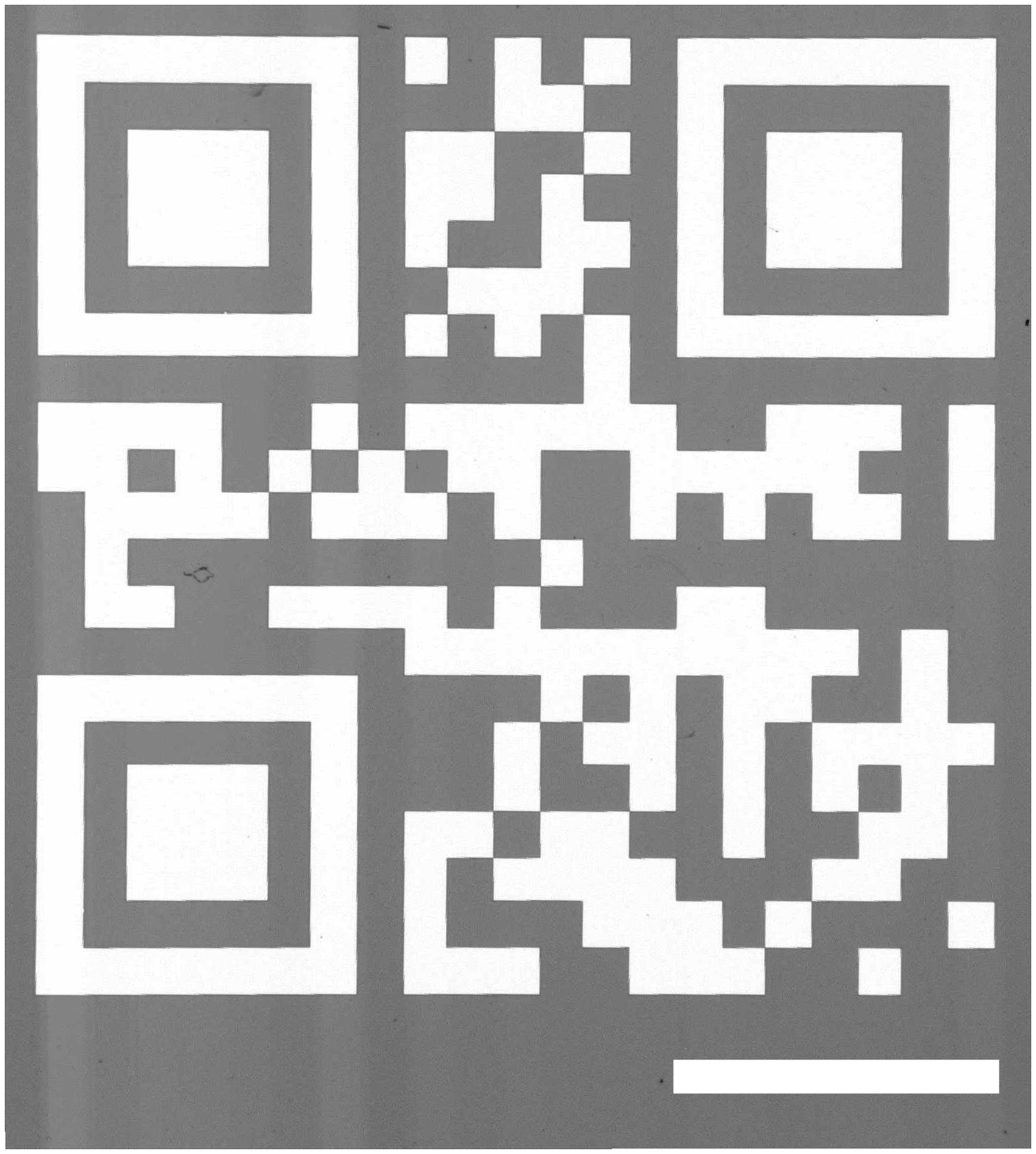
   }
	\hskip0.01mm
	\subfigure{
	\def\svgwidth{0.25\columnwidth}
  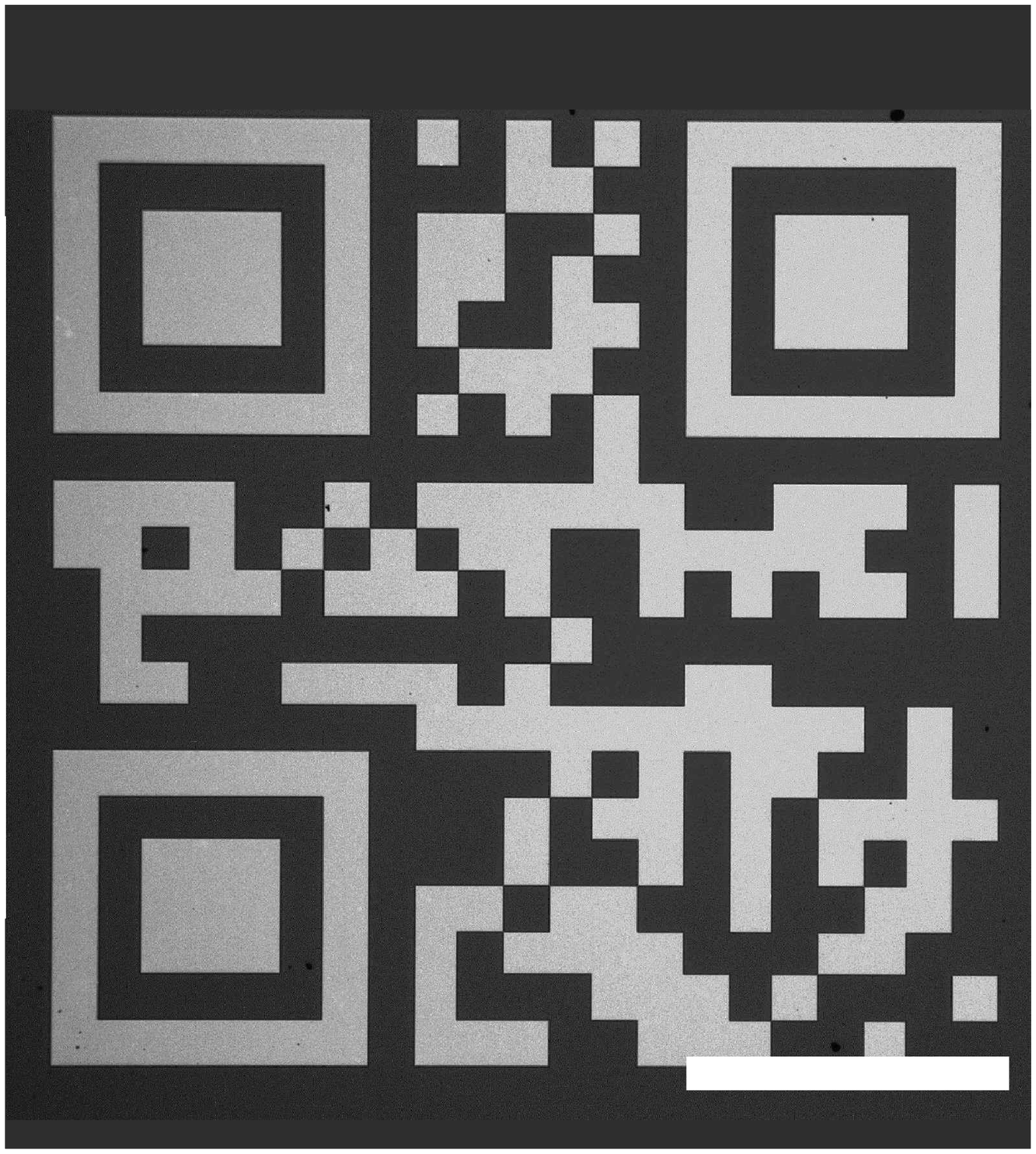
   }
 	\hskip0.1mm
	\subfigure{
	\def\svgwidth{0.25\columnwidth}
  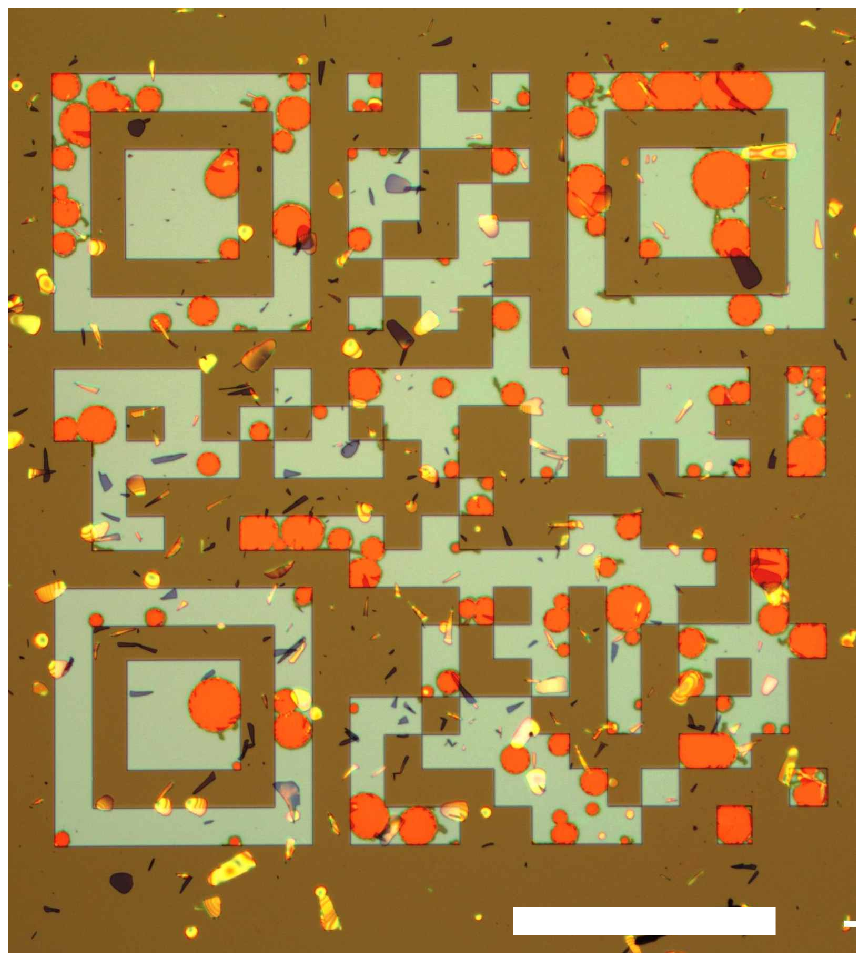
   }

  \caption{Optical microscope images of the same 
  QR code left: after fabrication, center: after 2 hours at \SI{613}{\kelvin} and right: 2 hours at \SI{763}{\kelvin}}
  \label{fig:QRcodes}
\end{figure}

The sample with the QR codes was exposed to 
temperatures of \SI{513(5)}{\kelvin}, \SI{613(5)}{\kelvin}
and \SI{713(5)}{\kelvin}. Each temperature increase causes
a reduction in the number of readable QR codes by the decoding 
algorithm. This is caused by cracking of the top $\text{Si}_3\text{N}_4$ as can be seen in figure \ref{fig:QRcodes}.
The unreadable QR codes are not visibly damaged and the tungsten is still present.

The misreading of the information is caused by the readout
using an optical microscope without a monochromatic light source. 
The images are taken using a top mounted camera and contain a multitude
of colours, caused by the variation in $\text{Si}_3\text{N}_4$
thickness due to the cracking. The very simple detection software 
was unable to correctly assign a black
or white colour to multitude of colours caused by the
cracking of the top siliconnitride layer. 

\begin{figure}[t]
  \centering
  \subfigure{
  \def\svgwidth{0.45\columnwidth}
   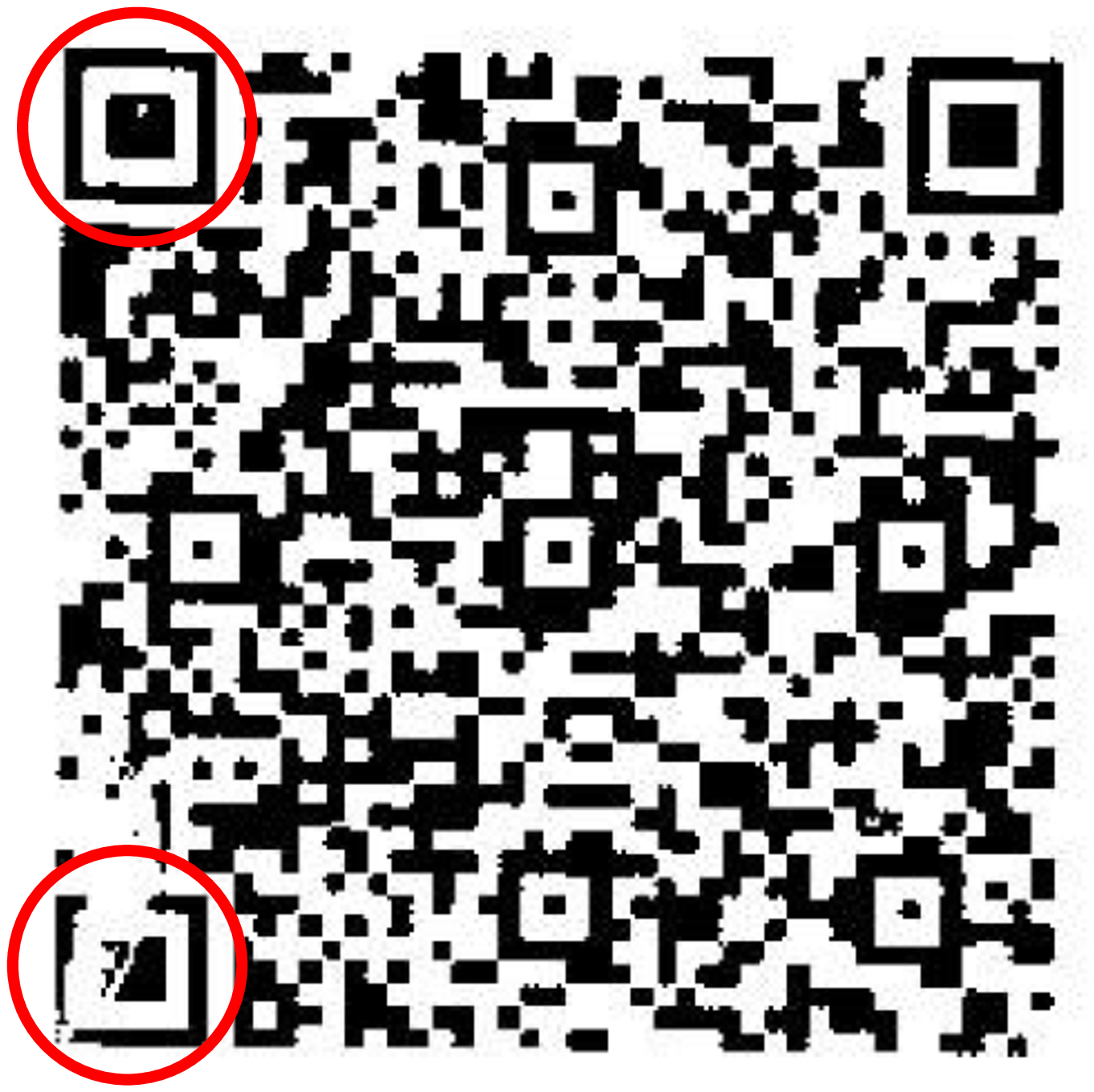
   }
	\hskip0.3mm
	\subfigure{
	\def\svgwidth{0.45\columnwidth}
  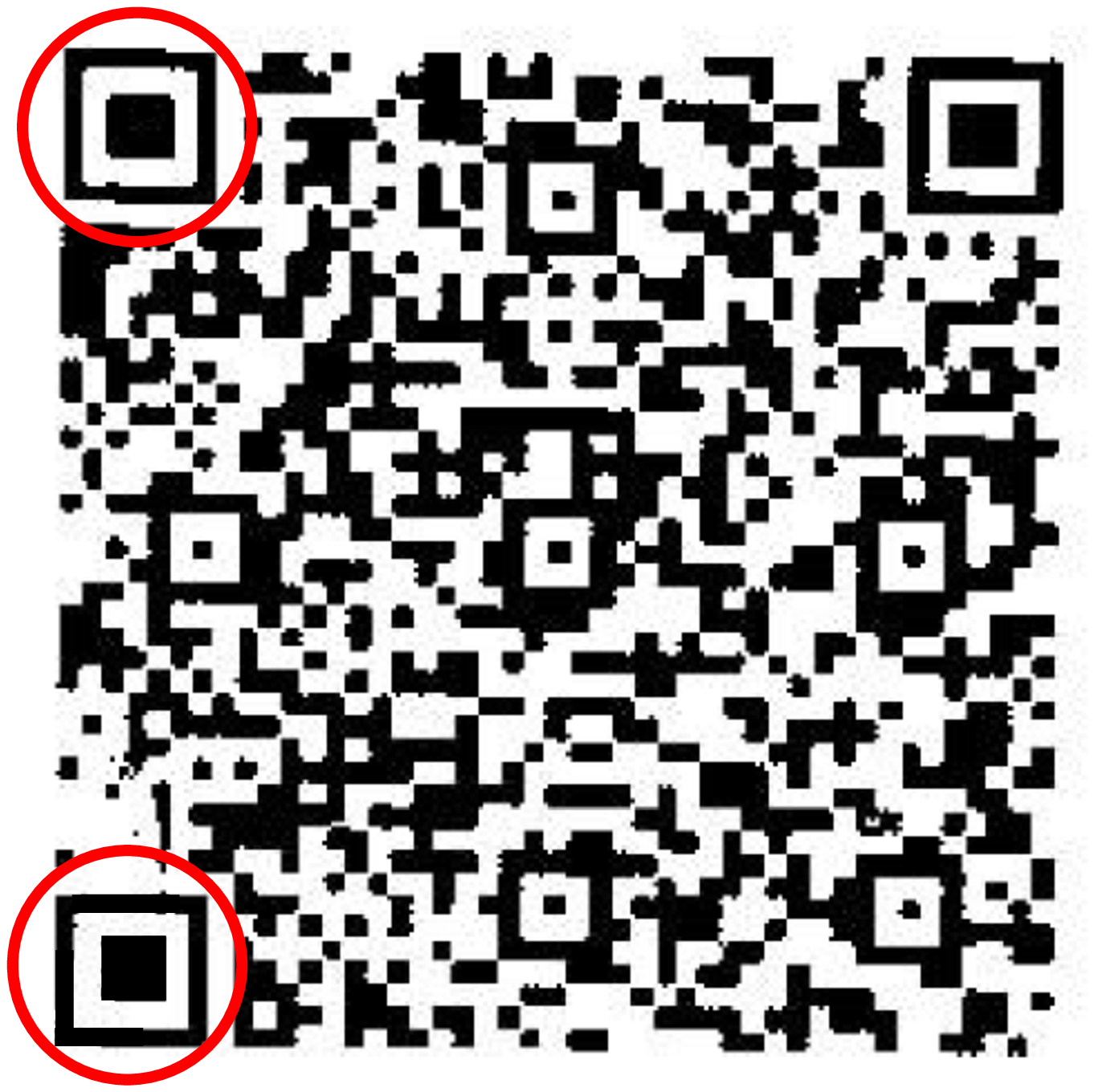
   }

  \caption{Manually repaired QR codes
  by fixing the finder patterns in the red circles.}
  \label{fig:QRRepaired}
\end{figure}

Due to the complexity of the QR code, damage to some areas 
affects the readability more than other areas~\citep{Vongpradhip2013}. When for instance the finder  patterns
are damaged, the QR code can not be read anymore. This
 can be seen from the example
in figure \ref{fig:QRRepaired} where the finder patterns of a damaged QR code
are manually repaired and the QR code becomes readable again.

A single wavelength microscope or more advanced detection software 
might solve this problem for the QR codes.

\subsection{Line patterns}
Table \ref{tab:testingtemperatures} shows that an ageing test at
\SI{445}{K} for 1 hour is sufficient to prove that the line sample would
survive for at least 1 million years. The sample was kept in
an oven at \SI{473}(5){K}
for approximately
one hour. Figure \ref{fig:SEMimages}~c and d show SEM images of the sample
before and after the test. We observe no visible degradation of the
sample, which indicates that this
sample would still be error free after 1 million years. Lower temperatures
have not been tested because the PECVD nitride is deposited at
temperatures of $\sim$\SI{573}{\kelvin}. This means that below this temperature 
damage would have occured during deposition. 
The sample has furthermore been tested at \SI{723(5)}{\kelvin} 
and \SI{848(5)}{\kelvin} without visible damage to the sample.

A test temperature of \SI{1373(5)}{K} for four hours, which can
occur in a hot spot during a house fire, showed that the sample
was completely destroyed and the tungsten lines could not be recognised 
anymore. The $\text{Si}_3\text{N}_4$ seems to have peeled off
probably due to variations in thermal expansion coefficients of
the top and bottom layer $\text{Si}_3\text{N}_4$ and the W.

\begin{figure}[t]
  \centering
  \def\svgwidth{0.55\columnwidth}
  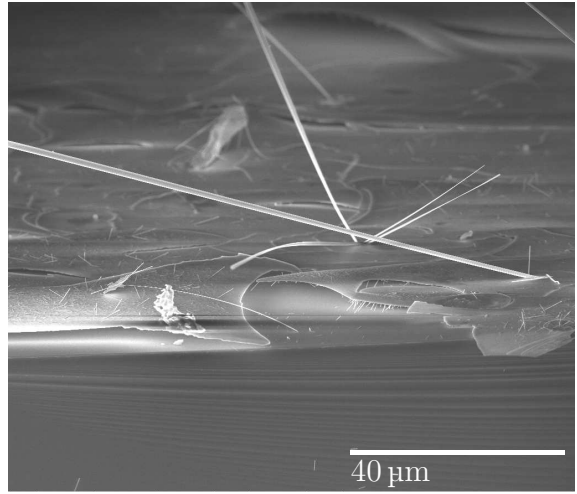
  \caption{SEM image of the whiskers growing from the top of the sample}
  \label{fig:Whiskers}
\end{figure}

\begin{figure}[ht]
  \centering
   \def\svgwidth{0.65\columnwidth}
   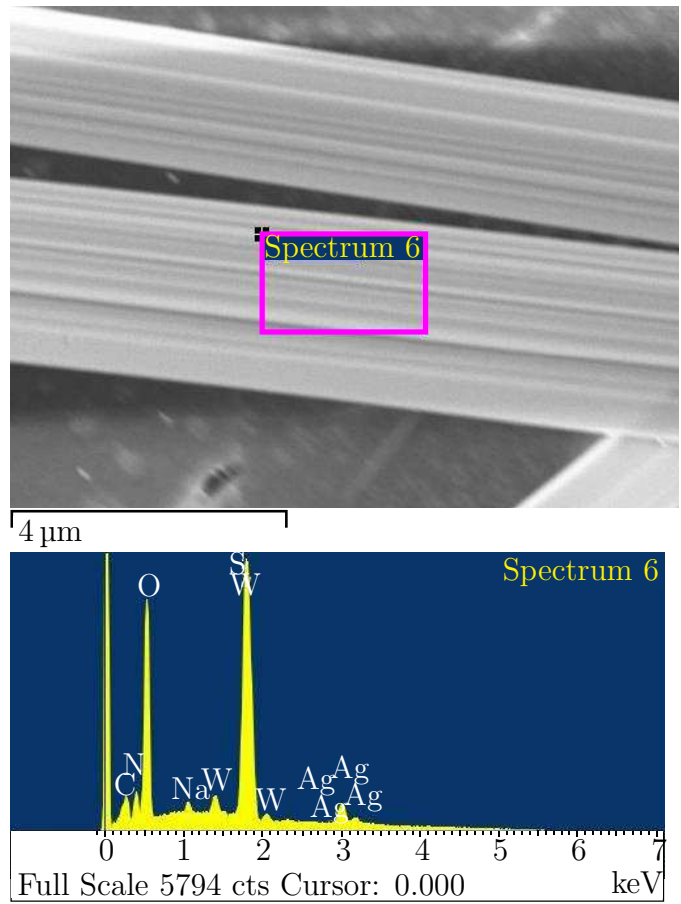   
  \caption{Top: SEM image of the whisker. Bottom: EDX information of the whisker
  showing large contents of W and O}
  \label{fig:WhiskersEDX}
\end{figure}

After exposing the sample to a temperature of \SI{923(5)}{K} for 4 hours 
we observed ``whiskers'' growing from the top of the sample as can
be seen in figure \ref{fig:Whiskers}. Here we also see the 
peeling of the $\text{Si}_3\text{N}_4$ layer. 
After inspection using an SEM with EDX capabilities we found
that the whiskers contain high levels of tungsten and oxygen as
can be seen in figure \ref{fig:WhiskersEDX}. What likely 
has happened is that the top $\text{Si}_3\text{N}_4$ deposited
by the PECVD process starts to exhibit cracks and oxygen can interact with
the tungsten. Under the influence of oxygen and the high temperature, WO whiskers are formed, similar
to the ones described by 
\citet{Cho2004_WO}.

Higher temperatures should be possible when the thermal 
expansion coefficients of the top and bottom layers are 
matched with the metal layer and the lines are replaced by islands. Previously
it was already found that a sample with small dots 
embedded in $\text{Si}_3\text{N}_4$ was able to survive 
temperatures up to \SI{1073}{K} for 30 minutes without 
degradation~\citep{Pei2009}. Whether a sample with 
larger islands would also survive these temperatures 
remains to be investigated.

\section{Discussion}

The initial attempt to create a medium containing embedded data
which is able to survive for 1 million years is promising. 
The optical readable data in the form of QR codes was 
able to survive the temperature up to \SI{713(5)}{\kelvin}.
The amount of readable QR codes decreases at higher 
temperatures, but this seems to be largely due to the 
detection scheme as the tungsten is still present. 
The medium can survive high temperatures, up to 1 hour 
at \SI{848(5)}{\kelvin}, without visible degradation of the medium but at 
higher temperatures the medium degrades rapidly. 

At lower temperatures which corresponds to a storage time of 
1 million years or more, the data carrier survives. 
If we only take the Arrhenius law into account this should be 
sufficient to prove that data will survive for 1 million years. 

It is likely however that the Arrhenius law that we
use is too simple to describe the real ageing process. If the energy
barrier consists of intermediate steps, a cascaded Arrhenius law would
be required and data deterioration could occur much faster than
expected from these results~\citep{Bardou2000,Bertin2008}. 

We do believe however that diffusion of the tungsten is not the
primary concern as shown by the cracks caused by the elevated 
temperature test. Prolonged exposure of the tungsten to oxygen could
lead to the creation of the whiskers. Furthermore low attempt
frequency processes like erosion, fracturing and vandalism might have
a much larger influence on the lifetime of the disk than diffusion.

The suggested medium is also interesting for fire-proof archiving. Possible solutions 
to ensure that information could also survive
 temperatures in the excess of \SIrange{973}{1173}{\kelvin} 
 would be to have a better encapsulation process
ensuring that the thermal expansion coefficients of top and bottom
layers are matched with the metal layer. However house fires can contain hot spots
with temperatures above \SI{1473}{\kelvin} which might still prove to be too
high for the medium to survive. The $\text{Si}_3\text{N}_4$ or other 
encapsulation material should be dense in order to limit the 
diffusion of oxygen. A different material acting as ``data''
could also solve the reaction with oxygen. Materials
with a slightly lower melting point than the \SI{3683}{\kelvin} of tungsten should be no problem.

When the sample survived a temperature of \SI{848}{\kelvin} 
for one hour we would, according to the theory, have proven that the 
sample would last \num{9E29} years, which is highly unlikely.


\section{Conclusion}

Initial calculations show that it is possible to store data for over
1 million years, or even 1 billion years, with reasonable energy
barriers in the order of 70 k$_\text{B}T$. To prove that the data will not
disappear over this time period, one can perform accelerated ageing tests at
moderately elevated temperatures (\SI{461}{\kelvin} for 1 hour to represent 1
million years).

A disk with data in the form of QR codes has been fabricated and was
able to survive the temperature tests, and therefore will survive one
million years of storage according to theory. Data readout failure at
even higher temperatures occurs because of the internal stress in the
layer stack at elevated temperatures, which leads to fractures in the
top layer, resulting in color changes in the readback image.
 
A model for a high density recording medium, consisting of W lines with 
a width below \SI{100}{\nano\meter} embedded in a
Si$_3$N$_4$ matrix, has been successfully fabricated. An accelerated
ageing test was performed by storing the sample at \SI{473}{\kelvin} for one
hour. There was no visible degradation of the sample or the tungsten
lines. If we only take diffusion into account this proves that the
sample will survive for well over 1 million years when stored at \SI{300}{\kelvin}.



Exposure of the medium to higher temperatures shows degradation due to the 
difference in thermal expansion coefficients between the two different
types of Si$_3$N$_4$ and the W lines. The top layer of Si$_3$N$_4$
starts to shows cracks and the W is exposed to the environment. 
This leads to the ''whiskers'' being grown under the influence 
of oxygen and high temperature.

\section{Future work}

Reduction of the amount of decodable QR codes is likely caused by the
white light source of the microscope and the simple detection
scheme. Solving the problem with the variation in stress between the
nitride and the tungsten layers would probably also solve the readout
issues of the QR codes.

Another influence on the sample could be etching
of the nitride due to acids. There are known acids
like phosphoric acid which can etch siliconnitride.
Also erosion due to influences of for instance wind
and sand can deteriorate the medium. 
A suitable container or storage location
can decrease these influences. To properly investigate
these influences more advanced tests will have to 
be performed on the medium.

\section{Acknowledgements}
The authors would like to thank Kechun Ma and
Johnny Sanderink for helping with 
the fabrication as well as Mark Smithers for the
SEM and EDX measurement.





\bibliographystyle{unsrtnat}








\end{document}